\documentclass[12pt,preprint]{aastex}

\usepackage{natbib}

\slugcomment{Submitted to the Astrophysical Journal}

\shorttitle{Doppler-Shift Measurement of Meridional Circulation}
\shortauthors{Roger K. Ulrich}

\begin{document}

\title{Solar Meridional Circulation from Doppler Shifts of the {\sc Fe I} line at $\lambda 5250$\AA\ as Measured by the 150-foot Solar Tower Telescope at the Mt.\ Wilson Observatory}

\author{Roger K.\ Ulrich}
\affil{Department of Physics and Astronomy, University of California,
    Los Angeles, CA 90095-1562}

\begin{abstract}
Doppler shifts of the Fe I spectral line at $\lambda5250$\AA\ from the full solar disk obtained over the period 1986 to 2009 are analyzed to determine the circulation velocity of the solar surface along meridional planes.  Simultaneous measurements of the Zeeman splitting of this line are used to obtain measurements of the solar magnetic field that are used to select low field points and impose corrections for the magnetically induced Doppler shift.  The data utilized is from a new reduction that preserves the full spatial resolution of the original observations so that the circulation flow can be followed to latitudes of 80$^\circ\,$N/S.  The deduced meridional flow is shown to differ from the circulation velocities derived from magnetic pattern movements. A reversed circulation pattern is seen in polar regions for three successive solar minima.  An surge in circulation velocity at low latitudes is seen during the rising phases of cycles 22 and 23.
\end{abstract}


\section{Introduction}
The mechanisms producing sun's eleven-year cycle of solar activity remain only partially understood.  The magnetic manifestations of the cycle in the form of sunspots and related magnetized areas such as faculae and plages have been observed for much longer periods than is the case for the large-scale flows of matter which are present in the form of differential rotation, meridional circulation and the torsional oscillations.  A number of good reviews of the solar dynamo and processes associated with it are available: \citet{2003A&ARv..11..287O}, \citet{2002RSPTA.360.2741T}, \citet{2009SSRv..144...53W} and \citet{2010LRSP....7....3C}.  This paper deals primarily with the topic of the meridional circulation.  Recent progress in modelling the solar dynamo by \citet{2006ApJ...649..498D} and \citet{2010GeoRL..3714107D} has shown the importance of the meridional circulation pattern for understanding the duration of the solar cycle.

Meridional circulation can be measured either by using the Doppler shift of photospheric spectral lines or by tracking features on the solar surface.  If carried out successfully, the Doppler shift method has the advantage of giving the motion of the solar plasma itself whereas the feature tracking methods may be influenced by the drift of the features relative to the solar plasma.  However, the Doppler shift method requires knowledge of the limb shift function as well as the utilization of an instrument with stable spectral sampling.  Also, for single observations supergranulation velocities contribute to the average Doppler shift by an amount large compared to the meridional circulation even in the context of the largest spatial scales.  Issues facing the use of the Doppler shift method have been reviewed by \citet{1985SoPh..100..141S} and will be discussed in sections (\ref{observations}), (\ref{MagEffects}) and (\ref{LS}) below.  Most likely due to these complications, many published reports based on the Doppler shift method have yielded inconsistent results (see the above review as well as discussions by \citet{1996ApJ...460.1027H} and \citet{1996SoPh..163...21S}).  However, the problems associated with the Doppler shift method can be mitigated and several reported results will be used and compared here: \citet{1979SoPh...63....3D,1996Sci...272.1306H} and the series prepared at MWO -- \citet{1988SoPh..117..291U,1993ist..proc...25U,2005ApJ...620L.123U}.  The feature tracking methods in the past most commonly used magnetic fields \citep{1993SoPh..147..207K,1994SoPh..149..231L,1996SoPh..163...21S,2010Sci...327.1350H} or sunspots \citep{1986ApJ...307..389H} while recently the methods of helioseismology use acoustic waves propogating or standing in the solar plasma \citep{2002ApJ...570..855H} as well as supergranulation as detected by wave patterns in the divergence of the solar velocity field \citep{2008SoPh..251..241G}.  The work by \citet{1996SoPh..163...21S} contains an extended discussion of the issue of drift between the magnetic fields and the underlying plasma.

This paper describes a new approach to the reduction of the MWO data set that retains the full spatial resolution of the original observations.  Previous reductions have relied on what is referred to as the Interactive Data Reduction or IDR data set.  A variety of data products can be extracted from the IDR data set but none have better spatial resolution than the $34\times34$ arrays that were used for example by \citet{1996SoPh..163...21S} and \citet{2005ApJ...620L.123U}.  Section (\ref{observations}) describes the new reduction along with a summary of the instrumentation and its impact on the issues related to Doppler shift velocities.  Section (\ref{MagEffects}) discusses the important problem of the magnetically induced apparent velocities and the method of mitigating this effect.  Section (\ref{LS}) provides a summary of the treatment of the limb shift.  The method of extracting the meridional circulation function and the primary new results are presented in section (\ref{results}).  The dependence of the meridional circulation on the solar cycle is shown clearly in this section.  The relationship between meridional circulation results based on the Doppler shift approach and the magnetic pattern tracking approach is discussed in section (\ref{comparisons}).  Conclusions are presented in the final section.

\section{The Magnetic and Doppler Data Sets}
\label{observations}

The observations that form the basis of the present study have been obtained using the facilities of the 150-foot solar tower telescope located at the Mt.\ Wilson Observatory (MWO).  Although solar records from MWO begin in 1906 and exist now in various formats, this study is restricted to the period 1986 to November 2009.  For the period 1982 to 1986 the optics of the exit slit is the same as used during the selected period; however, the number of observations per day was only one or two so that observations from this period would not pass our selection criterion.  In addition this period does not include a full solar cycle so that we would not be able compare cycle 21 to cycles 22 and 23.  For these reasons, the data from before 1986 has been left out of this study.  The data between 1967 and 1982 is in digital format and is useable for magnetic field studies and Doppler rotation studies but it is not useable for meridional circulation studies due to the lack of a stable spectroscopic band pass (see below in section \ref{LS}).  Observations each day consist of one or two full-disk scans using an entrance aperture of $12\arcsec\times12\arcsec$ designated as slow scans and zero to about twenty scans made through an entrance aperture of $20\arcsec\times20\arcsec$ designated as fast scans.  Prior to 1986 there are no fast scans available.  

The processing of the data from its raw form as recorded at the MWO 150-foot tower to a form that is easily useable currently requires the use of a very old VAX computer.  The time to process one year's observations through this system is approximately 2.5 days so that approximately a month of calendar time was required to process the data from solar cycle 22.  Conversion of the processing code to execute on a more modern computing system is currently being carried out but this task will not be completed for about a year.  The reductions considered in this report were carried out in 2006/2007 for cycle 23 and June to July 2010 for cycle 22.  The reduction code has evolved over the period from 1986 to 2005 but has been changed only in minor ways between 2005 and present.  

The raw array size produces images of the sun with $(N_x,N_y)$ pixels in (E/W,N/S) directions.  For the slow scans the dimension is $(N_x,N_y)=(200,145)$ and for the fast scans $(N_x,N_y)=(150,95)$.  The E/W dimension is greater than the solar angular diameter in arc-sec divided by 12 or 20 because the sampling is about 65\%\ of the pixel width in the $x$ direction.  The actual pixel locations are not in a regular $(x,y)$ grid due to a variety of control factors but the actual locations in both space and time are known with good precision and this information is used to interpolate each observed point onto a regular grid at a fixed time using the algorithms described by \citet{2006SoPh...235...17U}. This interpolation includes the correction for differential rotation.  The array size for the interpolated grid is $(N_x,N_y)=(256,256)$ for this study and the time the image is created for each day is 20:00 UT.  The range in time of observation is generally $\pm4$ hours from this time.  The variability in the location of the observed pixels relative to the final pixels is a form of dithering and results in an improvement of the final spatial resolution.  

Days were selected for inclusion in the analysis only if there were three or more observations.  Days with fewer observations tend to be under less than ideal conditions due to cloudiness or other factors.  Although days with more than ten observations have velocity images that clearly show the supergranulation structure while days with only three observations are dominated by the $5-$minute oscillations, the selection was left at three days in order to obtain better temporal sampling.

Subsequent to \citet{1988SoPh..117..291U} and prior to the report by \citet{2005ApJ...620L.123U} the analysis algorithm was improved to account for the center-to-limb dependence of the $\lambda5250.2$\AA\ line shape and to account for geometric irregularities in the location of successive image scan lines.  The center-to-limb dependence of the line shape is needed in the conversion of intensity imbalance between the blue and red sampling ports into a velocity shift because the slope of the line wings is not constant.  Figure (\ref{images}) shows a typical pair of images in velocity and magnetic field that result from the reduction process.  Note the magnetic features associated with Doppler image irregularities.  In particular, magnetic fields in the image center and in the near-limb regions appear to have an opposite effect on the Doppler velocity.  Our approach to accounting for the magnetic field effects is discussed in the next section.

\section{Sensitivity of Doppler velocities to Magnetic Field Strength}
\label{MagEffects}

It is evident from even a cursory examination of Doppler images of the solar disk that magnetized regions have velocities that are influenced by the fields.  At least three factors can produce this result: a direct thermal or dynamic process that causes a true change in the plasma motion, a modification in the granular scale convection that alters the effective limb shift of the region and a modification in the thermal and dynamical structure of the magnetic flux tubes so that the spectral line profile - especially its $C-$shape - is altered.  The first of these factors would be of interest to the overall mass flow pattern since it could contribute to the global circulation process.  However, since the remaining two factors cannot be ruled out, the approach adopted here is two-fold: 
\begin{enumerate}
\item All pixels with $|B|>20$ gauss are dropped.  The field strength used is that measured by $\lambda5250$\AA\ without saturation correction. 
\item Velocities from the nominally-flat Dopplergrams are correlated with the line-of-sight magnetic field $|B|$ to find a relationship
$$\delta v_{\rm mag}=\gamma_0+\gamma_1|B|\ .$$
\end{enumerate}

The overall zero point of the flattened Dopplergram is arbitrary for the MWO observations because no absolute standard of wavelength is available to the system.  The disk center velocity is taken to be this zero point since any large-scale and persistent surface velocities should lie on level surfaces with their vectors perpendicular to the line of sight at disk center.  However, the magnetic effect violates this assumption and correction for this effect produces a small offset compared to the original zero point; hence the term $\gamma_0$ in the above equation.  

The average of $|B|$ over all pixels in a fixed latitude range subject to the condition that $|B|<20\;$G gives a measure of the filling factor of weakly magnetized plasma relative to unmagnetized plasma.  In quiet areas the average absolute value is typcially $1.2\;$G while in active latitudes the average is $4.5\;$G or greater.  If the filling factor of magnetized plasma were 100\%\ we might expect the average to be near $10\;$G.  The actual values suggest that the active regions have a significant level of magnetism even away from sunspots while the quiet regions also have enough magnetization that it is important to correct for the effect.

Determination of an appropriate value for $\gamma_1$ is made difficult by the varying effect of individual active regions.  When specific daily average images are used to determine this coefficient, the result is quite unstable with values ranging between 0.0 and 0.8 m s$^{-1}$ G$^{-1}$.  The most stable result comes when all daily pixels from an entire year are used in the analysis.  The number of pixels from a whole year ranges between $1.0\times10^{7}$ and $1.4\times10^{7}$.  Scatter diagrams with points plotted on a plane defined by $|B|,v_{\rm los}$ could be used to illustrate the relationship between these two variables.  However, the large number of points appropriate to the present study cannot be shown in such a diagram since the plotted points would overlap even in areas of relatively low density.  Instead the distribution function is binned onto the 2-dimensional $(|B|,v_{\rm los})$ plane wherein each column is a velocity histogram of all pixels between a specified range in $|B|$.    

\begin{figure}
\epsscale{0.65}
\plotone{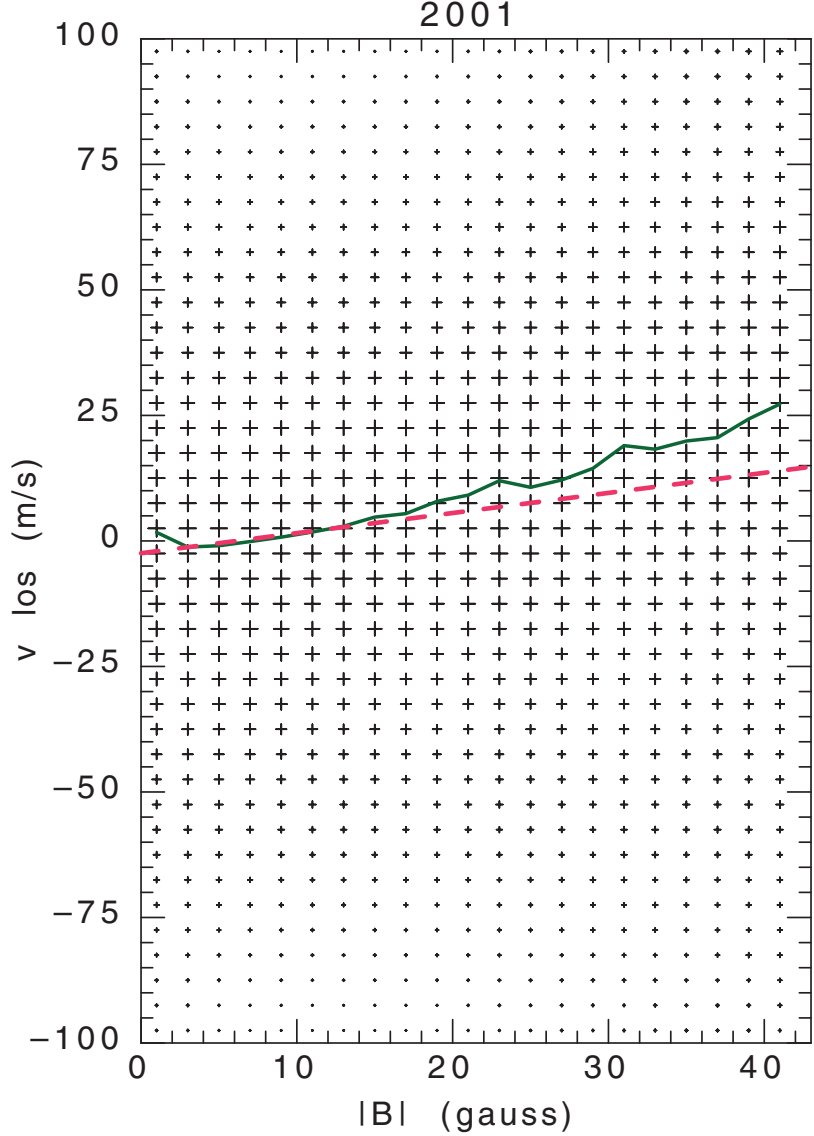}
\caption{A 2-dimensional histogram representation of the scatter diagram relating the apparent line-of-sight velocity and the absolute value of the magnetic field.  The plane of $(v_{\rm los},|B|)$ has been divided into cells and the number of pixels for the year falling inside each cell has been counted.  In the full plane there are between $1.0\times10^7$ and $1.5\times10^7$ points in each full year.  The numbers in the leftmost column have a maximum of $3\times10^5$ while the maximum in the rightmost column is only 600.  Consequently it is not possible to show a plotting symbol for each point on the scatter diagram.
Instead the size of each symbol is the ratio of the number in each cell divided by that column maximum.  The solid green line goes through the maxima of the histograms for each column and the dashed red line shows the adopted relationship between magnetic field strength and induced Doppler velocity shift.
\label{2DHisto}}
\end{figure}

\begin{figure}
\plotone{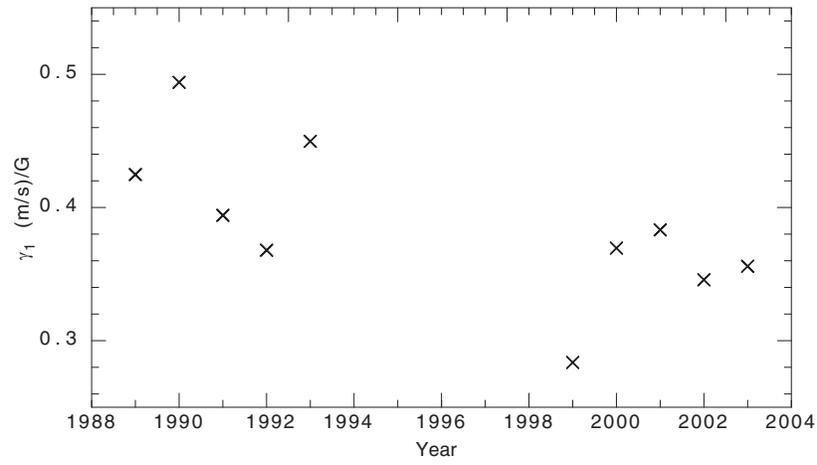}
\caption{The slope $\gamma_1$ relating the average velocity shift and absolute magnetic field.  Each value of $\gamma_1$ is determined from a linear fit to the average velocity within bins of absolute magnetic field averaged for whole years during periods of peak magnetic field activity.
\label{slopes}}
\end{figure}

The overall population distribution is shown in figure (\ref{2DHisto}) where each column is normalized by its peak value and the size of the plotting symbol is proportional to the resulting ratio.  The plotting symbol size was adjusted so that the symbols just do not overlap when the symbols are the largest.  The values of $v_{los}$ at the histogram maxima are shown as the green solid line.  Another indication of the relationship between $v_{\rm los}$ and $|B|$ comes from considering the average of $v_{\rm los}$ for the bins in $|B|$ instead of the maxima (the modes).  A fit to these averages for $|B|<20\;$G is shown in figure (\ref{2DHisto}) as the dashed red line.  Similar fits for periods of high solar activity yield the values of $\gamma_1$ shown in figure (\ref{slopes}).  For the remainder of this study a value for $\gamma_1$ of $0.4\,{\rm m}\;{\rm s}^{-1}\;{\rm G}^{-1}$ is adopted.  This quantity is uncertain by approximately $\pm0.05\,{\rm m}\;{\rm s}^{-1}\;{\rm G}^{-1}$.  The most important influence of this uncertainty comes in the active latitudes near 15$^\circ$ to 20$^\circ$ N/S where the average of the magnetic field strength for all pixels having $|B|<20\;$G is about $5\;$G.  Thus, the uncertainty from $\gamma_1$ is approximately $\pm0.25\,{\rm m}\;{\rm s}^{-1}$ -- too small to influence the discussion of the meridional circulation.

\section{Determination of the Limb Shift Function}
\label{LS}

The line-of-sight Doppler-shift velocity of the solar surface is known to depend on the center-to-limb angle $\rho$ in a manner that further depends sensitively on the spectral line used for the Doppler shift measurement and on the way the shape of this spectral line is sampled.  The resulting function is called the limb shift function and the determination of this function quantitatively requires a careful definition of what constitutes a limb shift rather than a large scale flow and a well-defined procedure to determine this quantity.  The limb shift function comes from small-scale convective elements that are hotter when matter moves upward through the photosphere and are cooler when matter moves downward through the photosphere as was first described by \citet{1978SoPh...58..243B}.  Although the shifts and asymmetries in spectral lines produced by convection can be modelled with a good degree of success \citep{2000A&A...359..729A}, when applied to full-disk, synoptic data such a detailed approach is inappropriate because of the extensive nature of the required application and because it is not clear whether all required model parameters can be constrained adequately in the context of a magnetograph velocity that comes from just two points on the spectral line.  Consequently, the MWO synoptic program has adopted the methods described below. 

\subsection{Method}
Our method of determining the limb shift function is that given by \citet{1988SoPh..117..291U}.  The essential steps are similar to the approach used by \citet{1979SoPh...63....3D} with some modifications appropriate to the higher spatial resolution MWO data.  The points selected for inclusion in the limb shift determination are all those whose $y$ coordinate value is between $\pm(4/17)R_\sun$ of $y=0$ where $(x,y)=(0,0)$ corresponds to disk center.  Furthermore, any points with $|B|>20\;$G were rejected along with a paired point on the opposite side of the central meridian.  The selected points were fit with a modified fourth order polynomial in $X=\left(1-\cos(\rho)\right)$ where the modification consists of dropping the term in $X^2$.  This function was adopted to be dependent on $\rho$ alone and subtracted from all line-of-sight velocities.  This fit has been applied to each observation with fitting coefficients derived for that observation.  Noise from non-circulatory solar velocity fields thus indirectly impacts the limb shift function for each observation and then influences the derived meridional circulation. The same non-circulatory velocities directly influence the meridional circulation as well so the combined effect will be larger than would be the case if the limb shift were known precisely. The discussion below shows that the random character of the non-circulatory velocity signals allows temporal averaging over a period of one month or more to reduce the combined direct and indirect errors to a level of less than 1 m$\,$s$^{-1}$.  The justification for the equatorial band fitting of the limb shift is discussed extensively by \citet{1988SoPh..117..291U} and those arguements have not changed.  That publication tentatively found the limb shift to be a sensitive function of the exact spectral band-pass sampling the blue and red wings of $\lambda5250$\AA. The variability in the limb shift function seen prior to the installation of fiber-optic image reformatters in 1982 has been followed by a very stable limb shift function confirming our tentative conclusion.  

\subsection{Equator-Crossing Flows}
Depending on the time of year, the sampled band is weighted more toward the northern or southern hemisphere.  An equator crossing flow like that reported by \citet{1998ApJ...504L.131S}, \citet{1999ApJ...527..967M}, \citet{2000PhDT.........9G} and \citet{2010Sci...327.1350H} would influence the limb shift function except for two factors: 
\begin{enumerate}
\item The analysis we use as described by \citet{1988SoPh..117..291U} includes a term nominally intended to remove spectrograph drift.  This term fitted and removed any linear trend across the full solar disk.  An equator crossing flow would be largely cancelled by this term.
\item The simplest interpretation of the equator crossing flow is that discussed by \citet{2005ApJ...621L.153B} wherein there is an error in the assumed position of the sun's axis of rotation.  For the case of the data from MDI discussed by \citet{2005ApJ...621L.153B}, this is a combination of uncertainty in the actual orientation of the spacecraft and a possible error in the Carrington ephemeris.  This effect is present in data products from MDI unless explicitly removed.
\end{enumerate}
As part of the study reported by \citet{2005ApJ...620L.123U}, an extensive search was made to identify the cause of the annual variations in the meridional circulation.  A mis-identification of the direction of the solar axis of rotation was one of the hypotheses considered at the time.  An artificial data set was created having this error and after reduction with the standard fitting functions, the effect of the wrong rotation axis was eliminated leaving no annual variations in meridional circulation.  A search for an equator-crossing flow using MWO data will require a different reduction algorithm that does not remove the full-disk velocity gradient.
Thus the meridional flow deduced from our method does not show an equator crossing flow but this fact does not establish that this flow is in fact absent.  For comparisons to other meridional flow results, any equator crossing flow in the non-MWO function is set to zero by adding a constant to the reported meridional flow functions.

\subsection{Time Dependence}
The observed solar surface velocity field includes solar contributions from oscillatory motions (the 5-minute oscillations), convective motions (supergranulation, granulation), meridional circulation and the limb shift as well as instrumental contributions such as spectral band-pass variations and scattered light.  In order to study the meridional circulation all other contributions must be removed from the signal.  The meridional circulation is long lived and global in scale.  Matter flows to high latitude where it sinks, flows back toward the equator where it encounters the stream from the other hemisphere and must rise. When the upwelling matter reaches the surface, it must spread horizontally since it cannot continue into space.  Due to the large scale of the flow it must be roughly anti-symmetric about the equator so that near the equator the direction of motion must be perpendicular to the equator.  When averaged over a year the line-of-sight is in the plane of the solar equator making the meridional flow near the equator also perpendicular to the line-of-sight.  These characteristics allow isolation of the meridional circulation: the long duration makes temporal averaging effective while the direction of flow means that near the equator only the limb shift can contribute to the average line-of-sight velocity.  The measurement of the solar surface velocity as part of the synoptic program at the 150-foot tower is done through a scanning aperture that is guided over the full solar surface with a temporal duration of 30 to 50 minutes. The oscillatory velocities are substantially undersampled relative to the dominant period of 5 minutes while the supergranulation convective velocity pattern may be correlated over a period of a day or so but on weekly or longer periods should be randomly distributed.  Both contributions are termed here ``non-circulatory'' and are assumed to have a zero mean relative to the limb shift plus the meridional circulation. 

The reduction algorithm carries out a fit of the limb shift function for each image and subtracts this function from the remaining line-of-sight velocities.  This approach limits the influence of variable spectral sampling and changes in scattered light from the telescope optics and from the atmosphere.  However, the regular determination of the limb shift allows changes in solar velocity fields in the equatorial band to be attributed to meridional circulation.  The non-circulatory velocities averaged over any specified area are random so they should have an amplitude that depends on $\rho$ but within a specified range in $\rho$ should be inversely proportional to the square root of that area.  The approach described below for the determination of the meridional circulation uses bands in latitude 2 degrees wide whereas the limb shift equatorial band is 26 degrees wide.  Thus the contribution of non-circulatory solar velocity fields to the error in meridional circulation only includes a fraction $(2/26)^{1/2}\approx 28$\%\ from the indirect influence of the limb shift function. 

\begin{figure}
\plotone{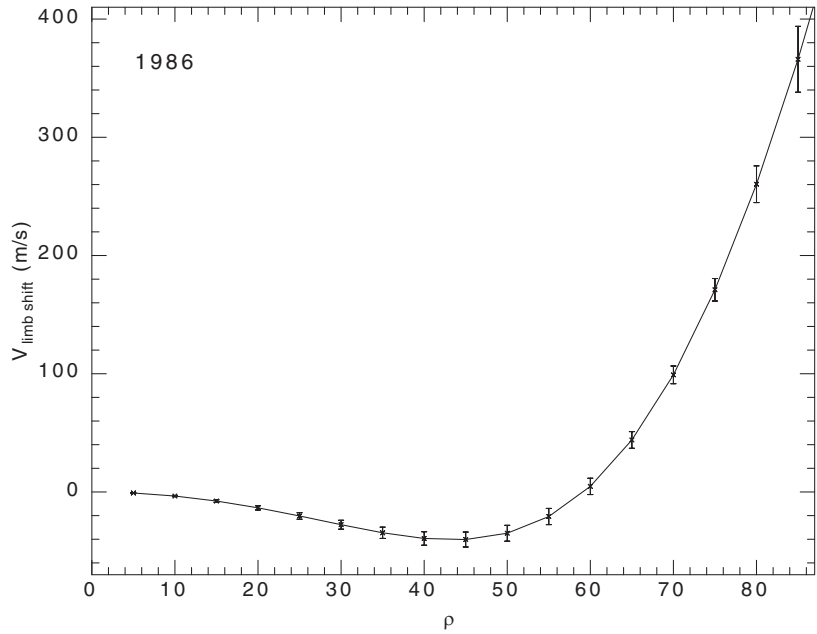}
\caption{This figure shows the limb shift function for $\lambda5250$\AA\ plotted against the center-to-limb angle $\rho$.  This function is the result of fits to every observation in 1986.  The value of the fit at $\rho=0$ was subtracted from the remaining fit values.  The average and standard deviation of the difference was calculated from the full set of observations with the averages being connected by the line on this figure and the standard deviation is represented by the error bars.
\label{center_to_limb}}
\end{figure}

The standard deviation of the limb shift function is large compared to the meridional circulation ranging from 7 m/s at $\rho=40^\circ$ to 25 m$\,$s$^{-1}$ at $\rho=80^\circ$.  Figure \ref{center_to_limb} shows the limb shift function and its standard deviation for the year 1986.  Other years are similar but have systematic differences from the case for 1986 due to changes in the optical configuration of the instrument.  Each year provides between 2500 to 3800 observations so as long as systematic effects are not present, temporal averaging should reduce the error due to the solar velocities to 2\%\ of the standard deviation -- an unimportant error over most of the solar disk but troublesome near the limb.  The method developed by \citet{1996ApJ...460.1027H} uses more of the solar disk to determine the limb shift function but the resulting function has contributions from both limb shift and meridional circulation.  These functions have a cross-talk matrix that allows the meridional circulation be be calculated from a preliminary limb shift function. Error propagation through this system may be difficult to evaluate, especially if high spatial resolution is to be obtained through extension of the analysis to high degree terms.

An important objective of the current study is to extract better spatial resolution than previously achieved with the MWO data.  Consequently, we checked the adequacy of the fourth order polynomial in the parameter $X$  in representing the averaged limb shift functions.  This check revealed a systematic ripple in the actual limb shift velocities compared to the fitted function.  A 6th order polynomial proved adequate to remove this systematic effect and a correction based on this polynomial was applied to the final velocities.  Average deviations from the fitted 6th order polynomial are shown in Figure \ref{ls_of_time} along with the difference between a 4th order fit and the 6th order fit shown as the dashed line.

\section{The Meridional Circulation}
\label{results}
\subsection{Method}
The observations that are the basis for this study are archived on dvd's and private on-line storage as raw records that include setup parameters and all measured quantities at each sampled point in the full-disk scans.  In this form the records are not easily useable because the sampled points are not on a rectangular grid and refer to a range of times within a 30 or 55 minute period of observation.  They are also in a digital organization that is accessible only to an older VAX 11/750 computer.  A preliminary step requiring approximately 1.4 minutes per observation is required to put the observed data into a form that can be accessed by modern computers.  This conversion has been carried out for those observations using $\lambda5250$\AA\ line between the times 1986 to 2009.   The resulting images were collected and reinterpolated into individual day fits format files of dimension $256\times256$ pixels and shifted by differential rotation to a standard time of 20 UT \citep{2006SoPh...235...17U}.  The images were then saved as data pairs $v_{\rm los},B$ with $v_{\rm los}$ in m$\,$s$^{-1}$ and $B$ in gauss.  Only days with three or more observations were included.

Our strategy for determination of the meridional circulation differs from that followed by \citet{2005ApJ...620L.123U} in that we have not attempted to track portions of the solar surface as they rotate through the visible hemisphere.  Instead, we take all points along lines of constant solar latitude and consider only the axially symmetric portion of the velocity deviations.  This will be sensitive to long-lived waves of the form discussed by \citet{2001ApJ...560..466U} but for long-term averages, the phase of the non-axisymmetric waves will be random and will not contribute to the determination of the meridional circulation which is axisymmetric.  Although the treatment of the time dependence in this study does not allow for the long-lived waves, it does share with \citet{2005ApJ...620L.123U} the ability to project the solar velocities onto planes in the E/W direction (zonal projection) and the N/S direction (sectoral projection).  The sectoral projection yields a vector in the meridional plane.  The direction of this vector relative to the local zenith is not constrained by the technique; however, our assumption that the motion is confined to level surfaces resolves this uncertainty.  It is nonetheless useful to retain a distinction between the sectoral velocities $v_s$ that are derived directly from the observations and the meridional velocities $v_m$ that use the level surface assumption.

\begin{figure}
\epsscale{.90}
\plotone{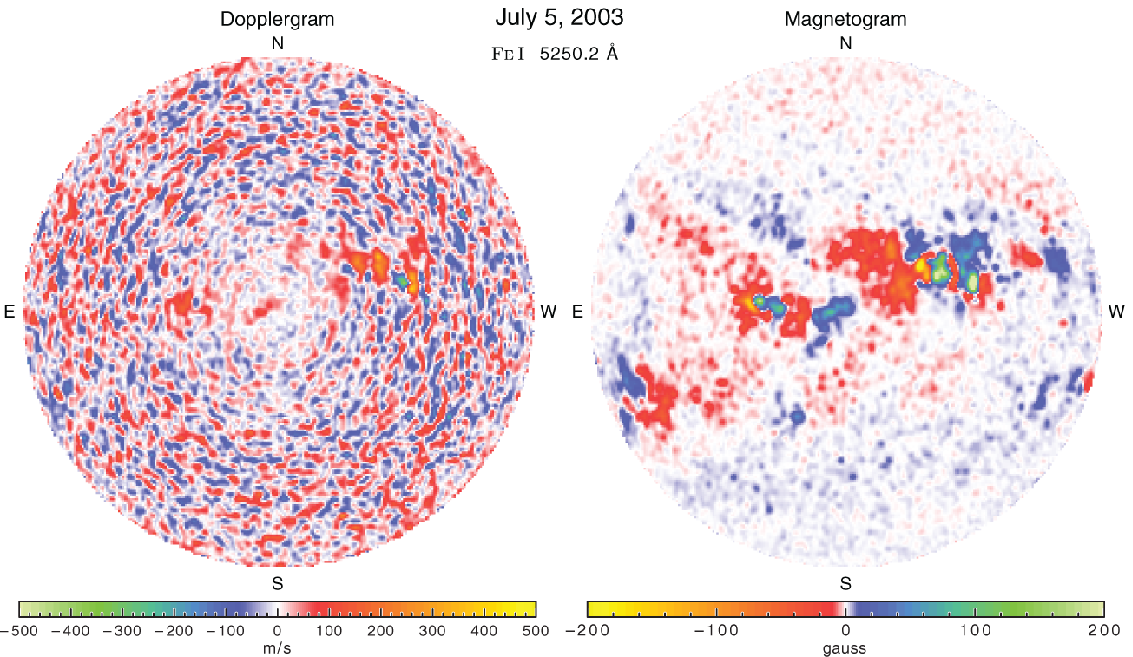}
\caption{Sample Dopplergram and magnetogram images from a typical day with significant magnetic activity.  Each is an average of all observations taken that day with each pixel shifted in position to correct for the differential rotation between the time the pixel is observed and a fixed time of 20:00 UT. This particular day included 13 observations with an entrance aperture of $20\arcsec\times20\arcsec$ and 2 observations with an entrance aperture of $12\arcsec\times12\arcsec$.  Note that the portions of the apparent disk near its center are occupied by several active regions, each of which is associated with excess positive line-of-sight apparent velocity.  Slightly south of the east limb is an active region associated with an excess negative line-of-sight apparent velicity and in the middle of the active region westnorthwest of disk center is a small area with a substantial negative apparent velocity.
\label{images}}
\end{figure}

\begin{figure}
\plotone{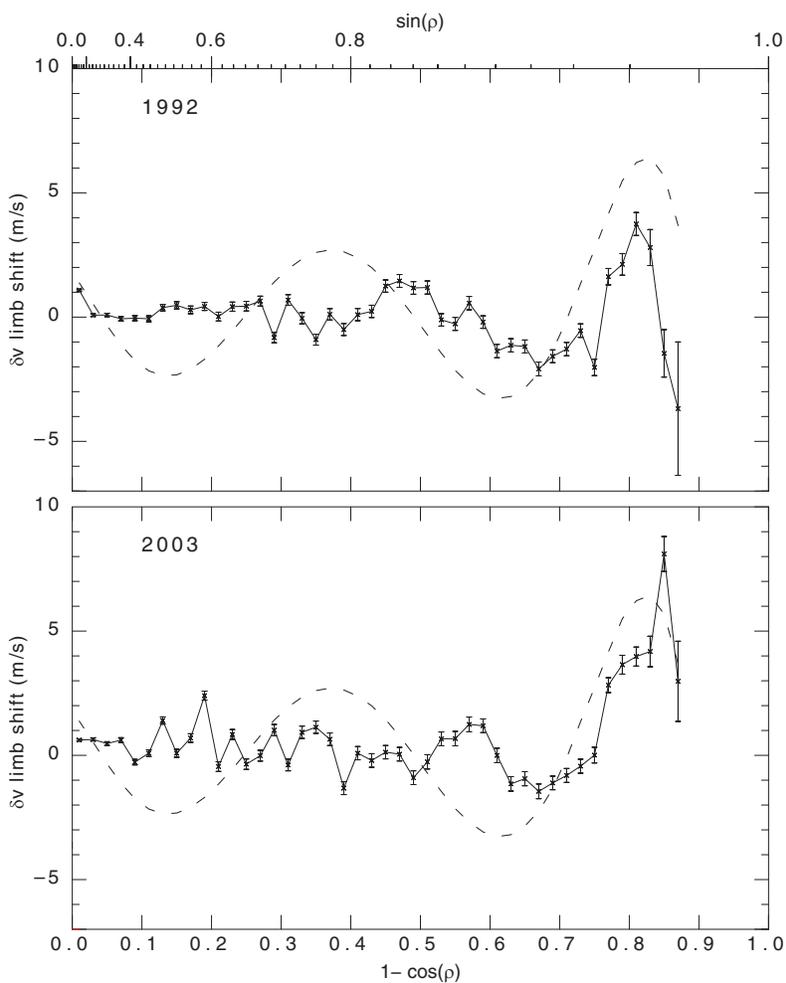}
\caption{Residual limb shift values averaged over whole years as a function of $1-\cos(\rho)$.  The points shown are calculated from the Doppler residual velocities like those shown in the Dopplergram of figure \ref{images} and have had both the fitted limb shift function for each image as well as the fixed 6th-order additional term already removed.  These average residuals are calculated in the same equatorial band as was used for the original limb shift fit with the averages being restricted to points with $|B|<20\,$G.  Two representative years are shown and the results for other years are similar.   The dashed line is the fixed 6th order term. \label{ls_of_time}}
\end{figure}

The raw observations such as those illustrated in Figure \ref{images} provide the line-of-sight velocity and the line-of-sight magnetic field on a rectangular grid $(x,y)$ where the $x$ coordinate is perpendicular to the sun's axis of rotation and the $y$ coordinate is parallel to that axis.  The origin for both axes is the center of the apparent solar disk, not the point where the solar equator crosses the central meridian.  The velocities shown in Figure \ref{images} have been corrected for a stationary meridional circulation pattern \citep{1988SoPh..117..291U,1993ist..proc...25U} in order to allow the variable details to be visible.  This static meridional circulation pattern was restored to the data by adding the latitude-dependent function back to the velocity deviations.  

The solar image is described quantitatively with the above rectangular $(x,y)$ grid when projected onto the sky and with heliographic latitude $b$ and central meridian angle $\delta L$ when projected onto the solar surface.
The $x=0$ line is the projection of the central meridian and the solar equator crosses this axis at a $y$ value of $y_{\rm eq}=R_0\sin(b_0)$ with $R_0$ being the solar radius and $b_0$ being the tilt angle of the sun's axis of rotation relative to the plane of the sky.  Along the central meridian the angle the level surface on the sun makes to the line of sight is $b-b_0$.  Near the center point an implied meridional flow along a level surface would be the line-of-sight velocity times $\csc(b-b_0)$ which could be very large if the line-of-sight velocity is dominated by noise at this point.  Consequently, special provisions are required to avoid contaminating the derived meridional flow with noise magnified by this mechanism.

The following steps were carried out for each selected day-average image:
\begin{enumerate}
\item Any pixel with $|B|>20\,$G was excluded.
\item Pixel positions $(x,y)$ were converted to solar coordinates $(b,\delta L)$.
\item The line-of-sight velocity at each pixel was corrected according to:
$$v_{\rm los}\longrightarrow v_{\rm los}-\gamma_1 |B|
$$
with $\gamma_1=0.4\;{\rm m}\;{s \rm}$.
\item The derived latitude values are assigned one of 256 index values: \[b\rightarrow j=\mathsf{Round}(128*\sin(b)+128.5)\]
\item Using an approach similar to that presented by \citet{2006SoPh...235...17U} the following sums were formed:
\[cv_j=\sum_{{\rm px}\in\; j} \cos(\delta L) v_{\rm los}\]
\[c\,c_j=\sum_{{\rm px}\in\; j} \cos^2(\delta L)\]
with the sums being carried over all selected pixels on that image.  This approach is based on the idea that the line-of-sight component of the meridional flow is proportional to $\cos(\delta L)$.  
\item A tentative average sectoral velocity is found from 
\[v_{{\rm s}\,j}={cv_j\over c\,c_j}\]
This result is only calculated for those values of $j$ with 15 or more pixels included in the sums.  Otherwise, $v_{{\rm s}\,j}$ is set to zero.
\item The region near disk center is smoothed with a $(0.25,0.5,0.25)$ filter (Denoting the smoothed quantity with an overline: $\overline{v_{{\rm s}j}} =0.25v_{{\rm s}j-1} + 0.5v_{{\rm s}j} + 0.25v_{{\rm s}j+1}$ ) to reduce the effect of noise on the conversion of a sectoral velocity into a meridional velocity.  This filter is applied once for latitudes $11.5^\circ<|b|<17^\circ$ and twice for latitudes $|b|<11.5^\circ$.
\item A quadratic polynomial for $v_{{\rm s}\,j}$ as a function of $b$ for $|b|<16^\circ$ is found by least square fitting. The polynomial minimum is defined as $v_0$.   
\item If $|v_0|< 5\;{\rm m}\,{\rm s}^{-1}$ the observation is accepted and $v_0$ is used to offset $v_s$, otherwise the observation is rejected and not used in the determination of the meridional flow. The band used for the quadratic fit is close to disk center and problems such as cloudiness and active regions near the fitted region can strongly perturb this solution leading to rejection under this test while days with good solutions typically have $-3\;{\rm m}\,{\rm s}^{-1}<v_0<2\;{\rm m}\,{\rm s}^{-1}$. Approximately 15\%\ of the otherwise acceptable observations are rejected by this test.
\item The meridional circulation velocity is found according to:
\begin{eqnarray*}
\hbox{If:$\quad|b-b_0|<0.28^\circ$\quad then\quad\quad}v_{\rm m}&=&0.0\\
\hbox{otherwise\quad\quad} v_{\rm m}&=&{v_{\rm s}-v_0\over\sin(b-b_0)}\ .
\end{eqnarray*}
\item All daily meridional circulation filess are collected into a single large data file for each year.  Each accepted day is treated equally with all others independent of the number of observations that went to form the file.  The values of $v_{\rm m}$ are rebinned into groups $2^\circ$ wide in latitude.  The mean and standard deviation of the values within each bin are calculated and used as the primary summary for each year.  For the solar cycle averages, the collection is expanded to include all years in the cycle prior to the binning and averaging step.  For some comparisons below with a different time interval the rebinning is managed as described here except that the boundaries of the collections are adjusted to agree with the comparison data sets temporal boundaries.
\end{enumerate}

\begin{table}\small
\caption{\small Comparison of the analysis of \citet{2005ApJ...620L.123U} ({\bf Old:}) to that used here ({\bf New:})\label{method_compare}}
\begin{description}
\item{\parbox{1.6in}{\bf Spatial Resolution:\hfill}}\quad  \parbox[t]{4.8in}{{\bf Old:} $34\times34$ array.\quad  {\bf New:} $256\times256$ array.}\vspace{-3pt}
\item{\parbox{1.6in}{\bf Limb Shift:\hfill}}\quad\parbox[t]{4.8in}{{\bf Old:} $|B|<20\;$G, 4th order polynomial in $\left(1-\cos(\rho)\right)$ fitted to each observation.\\[0pt]
{\bf New:} $|B|<20\;$G, fixed 6th order polynomial in $\left(1-\cos(\rho)\right)$ is added to daily fit.}
\item{\parbox{1.6in}{\bf Meridional Circ.\ :\hfill}}\quad  \parbox[t]{4.8in}{{\bf Old:} All points used, no magnetic correction.\\[0pt]
{\bf New:} $|B|<20\;$G, $\delta v_{\rm los}=-0.4|B|$ applied. }\vspace{-3pt}
\item{\parbox{1.6in}{\bf Image Zero Point:\hfill}}\quad \parbox[t]{4.8in}{{\bf Old:} Based on whole image without correction for the magnetic shift.\\[0pt] {\bf New:} Based on a quadratic fit of the line-of-sight component of the meridional circulation at disk center.}
\item{\parbox{1.6in}{\bf Annual Effect:}}\quad\parbox[t]{4.8in}{{\bf Old:} Detrended with superposed epoch analysis.\\[0pt]
{\bf New:} None detected in mid-latitudes. The polar and equatorial regions show unavoidable annual effects.  No detrending applied.} 
\end{description}
\end{table}

The present analysis is an extension of the work reported by \citet{2005ApJ...620L.123U}.  The principal differences between that work and this are summarized in Table \ref{method_compare}.  Two points are worth further discussion:

\begin{enumerate}
\item The effectiveness of the new reduction algorithms is indicated by the fact that the trends found by \citet{2005ApJ...620L.123U} in meridional circulation with a one-year periodicity are not present after the above steps have been applied.  A search for these effects through a monthly fit to the north and south meridional circulation velocities near $30^\circ$ revealed changes on the yearly time scale but each year was different and the variations are part of a longer term trend.  Latitude zones near the poles and near the equator do show strong changes within each year but these are a result of the basic geometry and cannot be avoided.  The step which is most responsible for eliminating the annual effect is that of correcting the zero point velocity.  Previously this number was determined from an overall fit to the velocities on the image.  Since the magnetic shifting effect was also not removed except for the limb shift determination, the shifting of the magnetic fields nearer and further from disk center introduced an annual variation in the zero point velocity and an associated annual variation in the meridional circulation.  

\item The zero point offset has an overall average value near $-1.5\,$m$\,$s$^{-1}$ with the consequence that the average value of the meridional circulation velocity increased by about 2$\,$m$\,$s$^{-1}$ relative to the result reported by \citet{2005ApJ...620L.123U}. 
\end{enumerate}

\subsection{Results}
The primary results of this paper are given in Figures \ref{yearly_circ} and \ref{cycle_22_23_compare}.  The time dependence of the meridional circulation pattern is striking in that the changes are large compared to the average values and that the changes occur smoothly from one year to the next even though observationally each year is completely independent of the others.  In addition, the similarity of the velocity profiles at similar phases of the solar cycle adds credibility to the interpretation of the patterns and their changes as being of solar origin.  

There are three periods near sunspot minimum 1986, 1996 and 2008.  All three have evident reversed flow cells near the poles.  Considerations of mass flow continuity in the solar atmosphere indicate that there should be a zone of subsidence where the flow converges and matter in the flow must leave the solar surface, settling toward the interior.  At sunspot minimum this zone of subsidence shifts away from the poles, perhaps due to a magnetic pressure effect from the polar fields which reach maximum strength at sunspot minimum.  The persistence of the reversed cell during the years 1986 to 1989 compared to the rapid disappearance after 1997 has been suggested to be an explanation for the greater length of solar cycle 23 compared to solar cycle 22 by \citet{2010GeoRL..3714107D}.  The solar cycle averages shown in Figure \ref{cycle_22_23_compare} confirm the significance of the difference between the two meridional circulation patterns.

Near the start of cycles 22 and 23 there is a strong surge in the strength of the meridional circulation near the equator.  Some of the details shown in Figure \ref{yearly_circ} may be influenced by the magnification of velocity pattern features due to the conversion of the sectoral velocity to the meridional velocity through division by a small number.  However, the larger surge near $20^\circ$ is not subject to that uncertainty.  There are stronger magnetic fields in this region compared to the solar minimum years but not compared to later periods of the solar cycle near sunspot maximum.  It is possible that this surge in meridional circulation is related to the onset of the new cycle when the magnetized gas rises from the interior.

The trend approaching the poles is not always consistent with the requirement that the meridional circulation must become zero at the singular point of the coordinate system.  This is especially true for the south pole in 1989, 1995, 1996, 1997 and 2008.  The south polar regions are more difficult to observe from Mt.\ Wilson due to the systematically poorer weather conditions during the winter to spring months when the southern hemisphere is preferentially visible.  It is also possible that the method of distinguishing between the sectoral and zonal flows through cosine weighted averaging is inadequate for the highest latitude regions.

Three of the years -- 1987, 1996, and 2001 -- indicate a possible equator-crossing flow of low amplitude.  In view of the inclusion in the fitting function of a term which removes any trend that is linear across the full image, it is unclear how the lack of symmetry across the equator could still be present for these years.  It is possible that this apparent flow is consequence of a lack of equatorial symmetry at higher latitudes which would induce an effect near the equator of opposite sign.  At this time it is doubtful that these features are of solar origin.

\section{Comparisons between Meridional Circulation Functions Deduced from various techniques}
\label{comparisons}
The various methods of measuring meridional circulation have yielded a variety of different flow amplitudes and latitudinal dependencies so that it is important to understand the nature of each type of observation.  Since the method used here is that of the Doppler shift, we start with this method.  The interaction between the meridional circulation and the limb-shift function requires that the spectral pass-bands used for the Doppler-shift determination be stable and well understood.  The individual daily analyses that are combined into the annual and solar cycle averages show a high degree of variability due to non-circulatory solar velocity fields so that it is essential to average over a large number of observations in order to obtain a stable result.  We have used temporal averaging whereas others such as \citet{1979SoPh...63....3D} through low spatial resolution and \citet{1996ApJ...460.1027H} through low order spherical harmonics have used spatial averaging.  Many of the early studies of the meridional flow based on the Doppler shift method (see the discussion and references in the introduction) suffer from inadequate averaging.  The modern velocity monitoring instruments that are designed to provide helioseismology data utilize a filter-gram method for determining the Doppler shift.  While these instruments yield good velocity changes, the determination of the zero point in such a system is challenging and there is no assurance that the filter pass-bands are independent of position on the solar image. The comparisons in this section between the meridional circulation functions derived from the MWO Doppler shift data and results from other methods are not intended as a complete survey of meridional circulation observations and are restricted to a few cases where the averaging is adequate and the spectral analysis (for Doppler-shift studies) is well controlled.

The methods considered are:
\begin{enumerate}
\item Direct Doppler-shift velocities.
\item Feature tracking of magnetic fields.
\item Feature tracking of acoustic waves through helioseismology.
\item Feature tracking of supergranulation through helioseismology.
\end{enumerate}
An important physical question is whether the underlying plasma and the features move together or have a relative drift as discussed recently by \citet{2010ApJ...271...xxxD}.  The magnetic fields in particular may respond to magnetic forces in addition to gas pressure gradients.  The acoustic waves extend below the solar surface so they represent a different set of matter than is observed at the solar surface so that comparison between Doppler-shift velocities and helioseismic velocities indicates the presence of a shear layer near the solar surface.  

\subsection{Comparison to Doppler-Shift Velocities}
\begin{figure}
\plotone{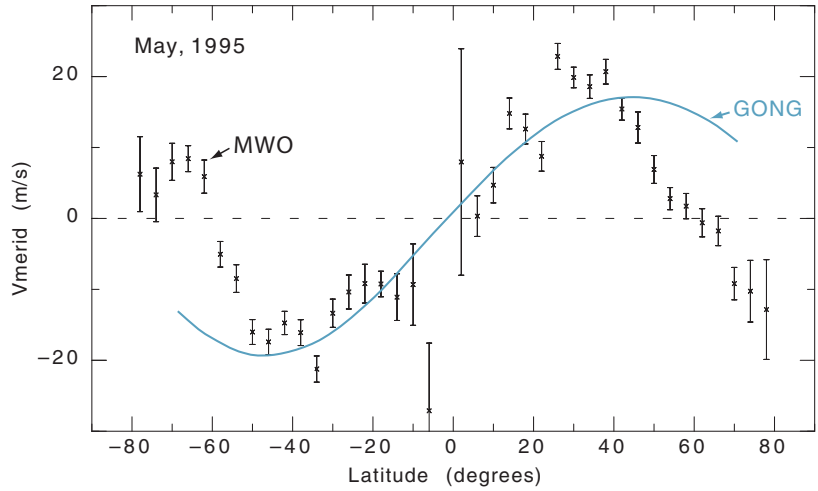}
\caption{This figure compares the Doppler shift result from the GONG instrument given by \citet{1996Sci...272.1306H} to the MWO meridional circulation.  The time period is just one month for both cases and uses much less averaging for MWO than for all other figures.  To avoid excess noise from the non-circulatory velocity fields, the bin size for this figure has been increased to $4^\circ$.  The erratic behavior near the equator is typical for a single month average.
\label{CompareToGONG}}
\end{figure}

This paper, \citet{2005ApJ...620L.123U} and \citet{1996Sci...272.1306H} use Doppler-shift velocities; \citet{1993SoPh..147..207K} and \citet{2010Sci...327.1350H} use magnetic feature tracking; and \citet{2008SoPh..251..241G} use feature tracking of the supergranulation velocity pattern.  There is consistency between results obtained using the same methods.  Figure \ref{CompareToGONG} shows the meridional flow deduced from a single month using the GONG system compared to the same month extracted from the observations discussed here.  The GONG reduction utilized the limb shift removal method of \citet{1996Sci...272.1306H} with spherical harmonics extending to 8th degree and found the solution for the meridional flow was dominated by the spherical harmonic of 2nd degree.  The analysis of MWO data for that month finds that the higher latitude pattern drops more quickly toward zero while at lower latitudes the flows match quite well in relative shape and amplitude.  The first observation of meridional flow by \citet{1979SoPh...63....3D} used a spectrographic instrument like that at MWO and found a meridional flow amplitude of 25 to 30$\;$m$\,$s$^{-1}$ which is 50\%\ larger than the MWO data shows for cycle 22.  In part because of the low spatial resolution of the system at the Wilcox Solar Observatory, Duvall's analysis did not include any correction or removal of areas with strong magnetic fields and this is a contributing factor for the higher flow rate he found.

\subsection{Comparison to Ring-Diagram Helioseismic Velocities}
\begin{figure}
\plotone{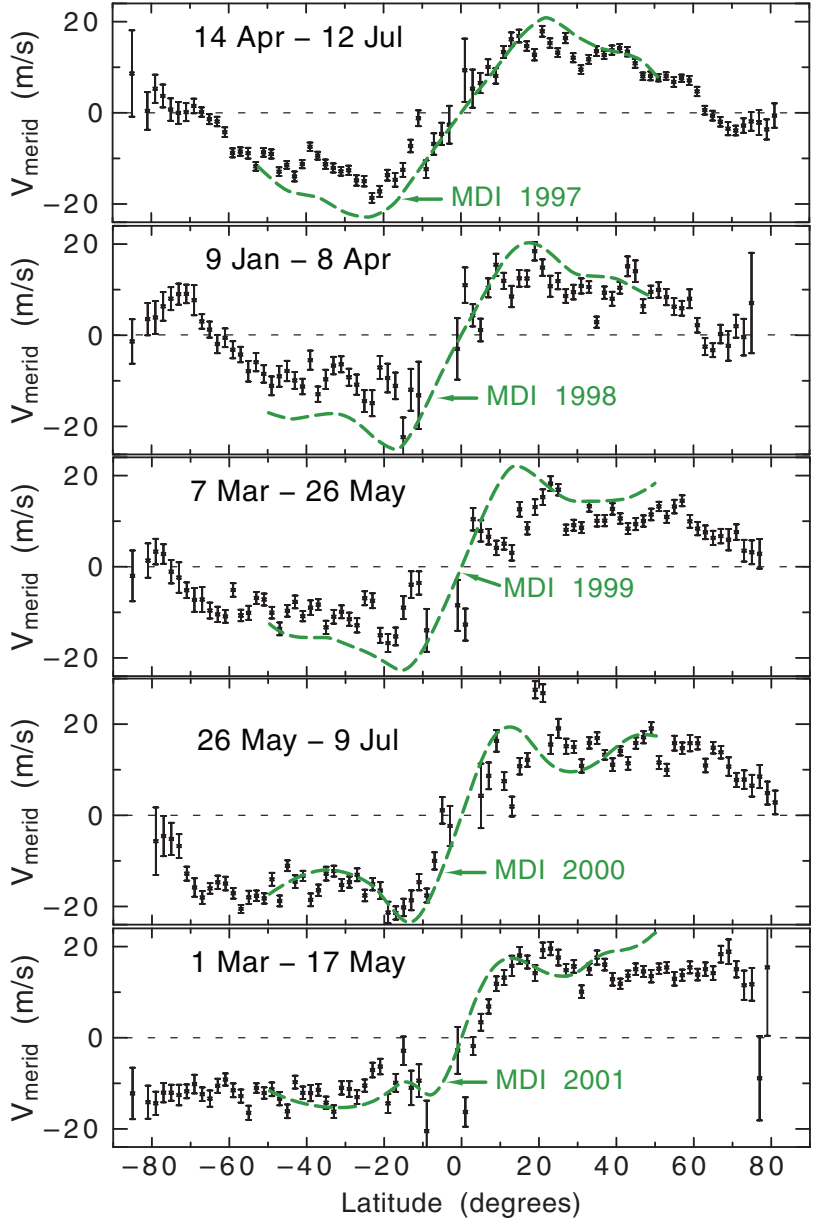}
\caption{This figure compares the meridional circulation derived from helioseismic analysis by \citet{2002ApJ...570..855H} using dense-pack ring diagrams to the MWO Doppler shift velocities.  The time intervals for the Doppler shift result were selected to be identical to those from the helioseismic results.  The black symbols with error bars are from the MWO analysis while the green dashed lines are from helioseismology.  The helioseismic meridional velocities have been shifted with a constant offset to yield curves that go through zero at the solar equator.  This condition is assurred for the MWO analysis and for the helioseismic velocities probably comes from the improperly determined angle between the SOHO spacecraft and the sun's axis of rotation as discussed by \citet{2005ApJ...621L.153B}.    
\label{CompareToRings}}
\end{figure}

Helioseismic feature tracking methods use acoustic waves to produce spatial variations that can be identified in image sequences.  The acoustic waves are not anchored by magnetic structures and should yield a result most closely associated with the Doppler-shift velocities.  The helioseismic approach that uses the acoustic waves allows for a determination of the velocity over a range of depths and for the comparison to the Doppler-shift results, the shallowest possible result from helioseismology is appropriate.  Figure \ref{CompareToRings} compares the meridional flows found with the MWO observations to those found by \citet{2002ApJ...570..855H} using dense-pack ring diagrams with data from the MDI instrument on SOHO.  For this comparison the time coverage of the MWO data was constrained to the time intervals provided by MDI.  The relatively small duty cycle for MDI comes from the telemetry restrictions that limit the full spatial resolution of the MDI instrument to just those periods referred to as the Dynamics Program when the Deep Space Network is available for one to two months per year.  Although the MDI data is also available for 1996, this year is omitted from figure \ref{CompareToRings} because the MWO system was in a transitional state for 4 months that year and the data reduction is non-standard.  Unfortunately, those months coincided with the period of the Dynamics Program.

The agreement between the Doppler-shift velocities and the ring-diagram velocities is very good both in amplitude of the flow and in the functional shape changes from one period to the next.  The helioseismic velocities are generally about 10\%\ greater than the Doppler-shift velocities.  This could be a consequence of the different depths of observation and a shear in the meridional circulation.  The active regions where a significant correction for the magnetic effect has been applied to the MWO velocities show the largest disagreement -- perhap indicating that the correction may be too large or that a similar correction should be applied to the MDI data.  The northern hemisphere comparisons are less satisfactory than for the southern hemisphere; possibly a consequence of the $b_0$ angle for these periods of observation which allows for a better view of the southern hemisphere.  At higher latitudes the only serious disagreement occurs in the north for 2001 where the MDI velocity increases sharply while the MWO velocity remains relatively constant.  Other than these features, the numerous points of agreement in the function shapes give confidence that solar variations are being observed by both systems.

\subsection{Comparison to Magnetic Feature Tracking Velocities}

Magnetic fields are found on the solar surface with a highly irregular spatial distribution.  Concentrations of magnetic field provide readily identifiable features that largely maintain their shapes but change their position on the visible solar disk.  The displacement of these features provides a very precise way of measuring a circulation velocity; however, as a number of authors have discussed \citep{1996SoPh..163...21S,2010ApJ...271...xxxD}, the solar plasma experiences forces due to gravity, rotation and pressure gradients whereas the magnetic fields experience these forces plus additional forces due to magnetic structure that may have a very large scale.  Thus the comparison between velocities derived from Doppler shifts to those derived from the drifts of magnetic features provides an opportunity to learn more about the forces experienced by the magnetized plasma.

\begin{figure}
\epsscale{0.8}
\plotone{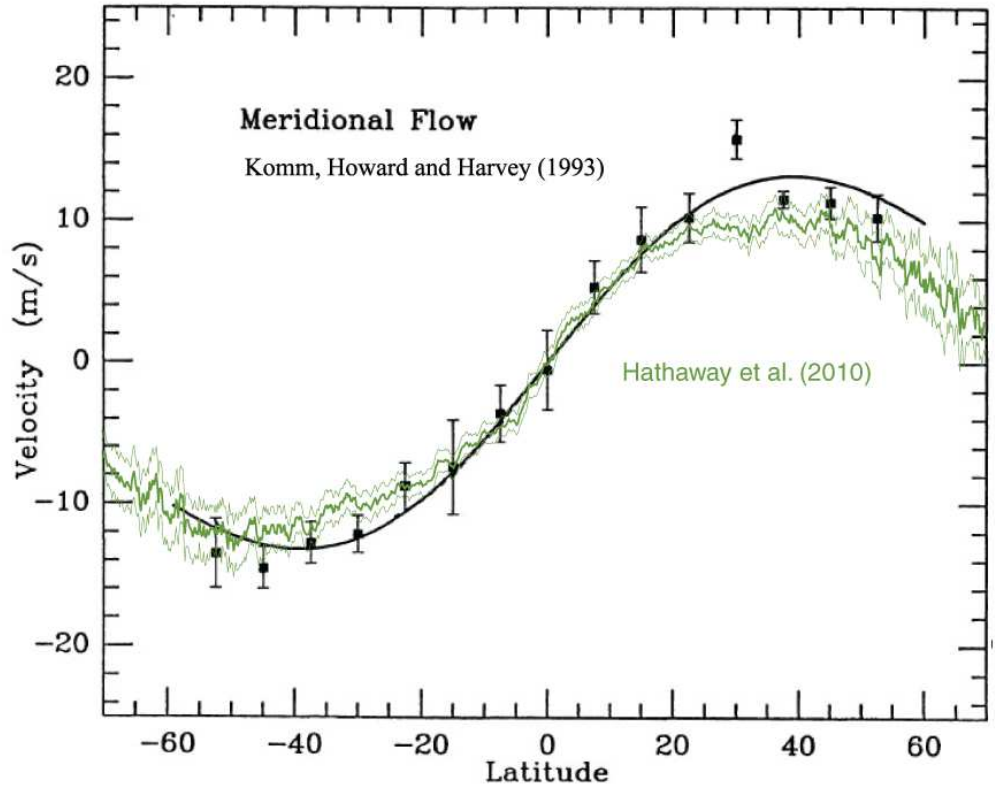}
\caption{This figure compares the meridional circulation derived from magnetic pattern tracking by \citet{1993SoPh..147..207K} to the more recent result presented by \citet{2010Sci...327.1350H} using the same technique. 
\label{MagneticFeatureTracking}}
\end{figure}

The work by \citet{2010Sci...327.1350H} is compared to the current analysis in figure \ref{CompareToMagFeature} for an average over cycle 23.  The significant differences in the meridional flow function shape and amplitude show clearly that these two methods are sensitive to different processes.  As further confirmation that the magnetic marker differs from the gas flow figure \ref{MagneticFeatureTracking} compares the results from \citet{2010Sci...327.1350H} to the earlier work by \citet{1993SoPh..147..207K} using the same method.  These two results agree very well with each other.

\subsection{Comparison to Supergranulation Tracking}

\citet{2003Natur.421...43G} showed that the convective pattern of the supergranulation could be tracked using a wave pattern in the divergence of the surface flow.  \citet{2008SoPh..251..241G} applied this method to data from MDI and gave meridional circulation results for the years 1996 to 2002.  The plots given in that paper have been transferred to the corresponding portions of figure \ref{yearly_circ} and are shown as the blue dashed lines in figure \ref{CompareToSG}.  Although the functional shapes are similar, the amplitude of the meridional flow velocity is much less than for the direct 
Doppler-shift velocities.  The fact that the shape of the curve agrees but the amplitude does not makes it difficult to interpret this tracking as being either related to the plasma flow rate or the magnetic feature rate.  It is too low to be the plasma rate and does not have the functional shape to be the magnetic feature rate.  The paper by \citet{2003Natur.421...43G} discussed the close connection between the supergranulation and magnetic features so it may not be surprising to find this result occupies an intermediate state between the plasma flow and the magnetic feature motions. 

\section{Discussion and Conclusions}
\label{DiscConcl}

This report is an extension of the work by \citet{2005ApJ...620L.123U} which included this velocity component along with the sectoral velocities (torsional oscillations) and two components of the magnetic field.  That work used summary data in the form of a $34\times34$ point arrays and did not include the effects of magnetic fields on the apparent velocity.  The present study utilizes $256\times256$ point arrays from a new reduction which provide much more detail in the meridional circulation velocity.  The impact of the magnetically-induced, apparent velocity shift is incorporated into the analysis.  A new velocity zero point has been determined from the minimum of a fit of the line-of-sight component of the meridional circulation on the solar central meridian to a quadratic function of $b-b_0$.  This new zero point has increased the derived meridional circulation velocities by about $2\;{\rm m}\,{\rm s}^{-1}$ and removes an annual effect previously found.

\begin{figure}
\plotone{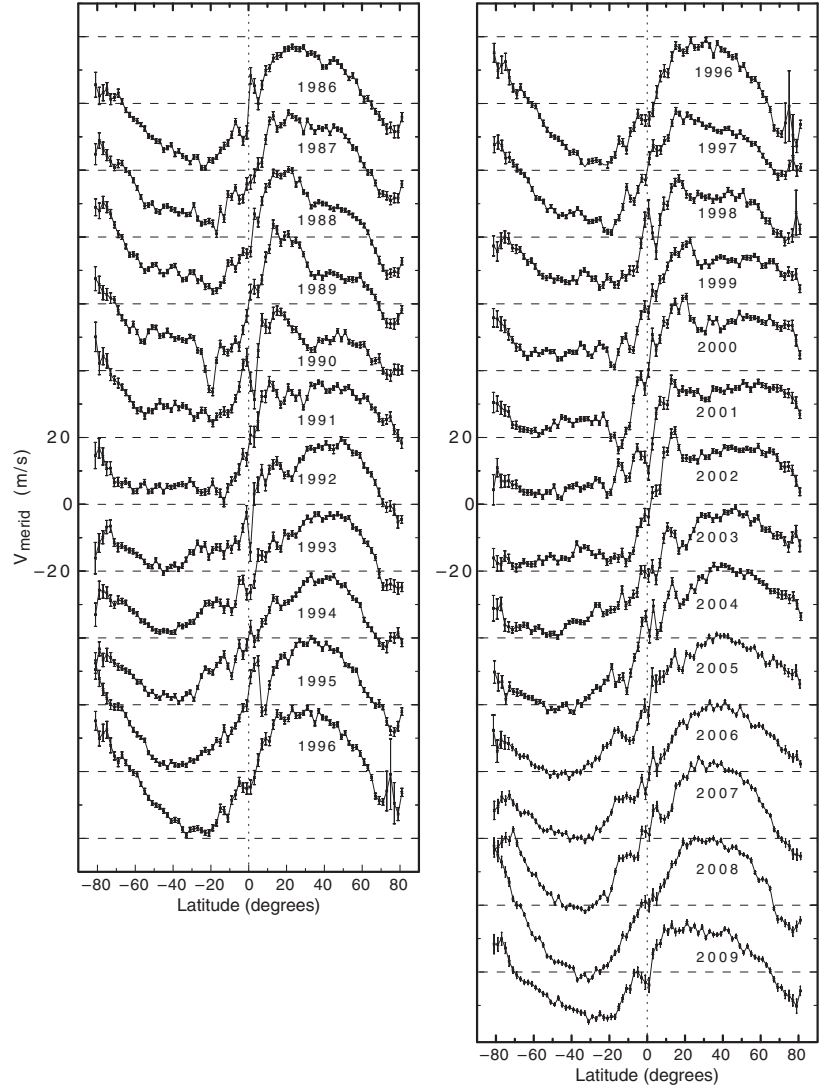}
\caption{Yearly results for the meridional circulation function.  Each day throughout each year is treated independently with the averages and standard deviations being calculated from the set of daily results.  The lines connect the average values  while the error bars give the error of the mean based on the number of days in the annual set.
\label{yearly_circ}}
\end{figure}

\begin{figure}
\plotone{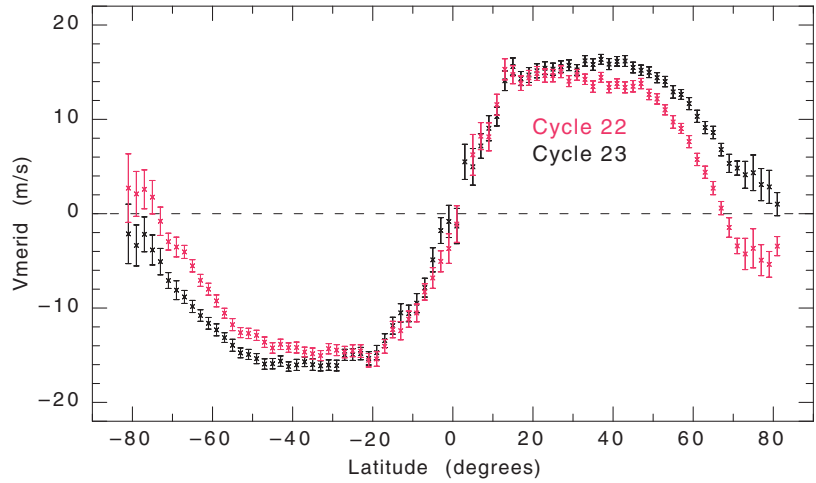}
\caption{Comparison between meridional circulation averages over the full cycles 22 and 23.  These averages and error bars are derived from the data shown in figure \ref{yearly_circ}.  The points are the averages over each solar cycle of the yearly averages.  The error in the cycle averages is assumed to be the average of the annual errors divided by the square root of the number of years in the cycle minus one.  The error bars shown here are three times this result.
\label{cycle_22_23_compare}}
\end{figure}

\begin{figure}
\plotone{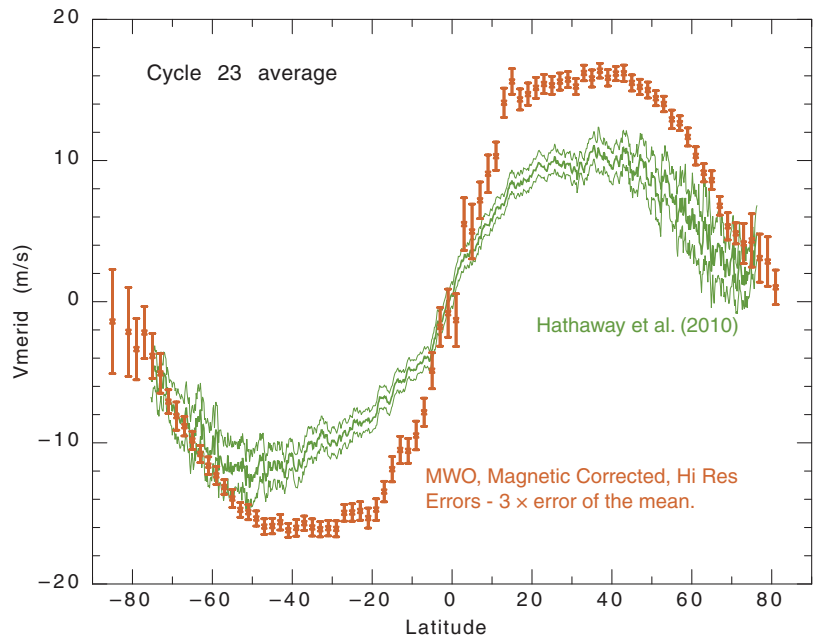}
\caption{Comparison between the Doppler velocity shift of the gas determined in this study witfh magnetic feature tracking velocities determined by \protect\citet{2010Sci...327.1350H}.  The MWO curve is the same as shown in figure \ref{cycle_22_23_compare}.
\label{CompareToMagFeature}}
\end{figure}

\begin{figure}
\plotone{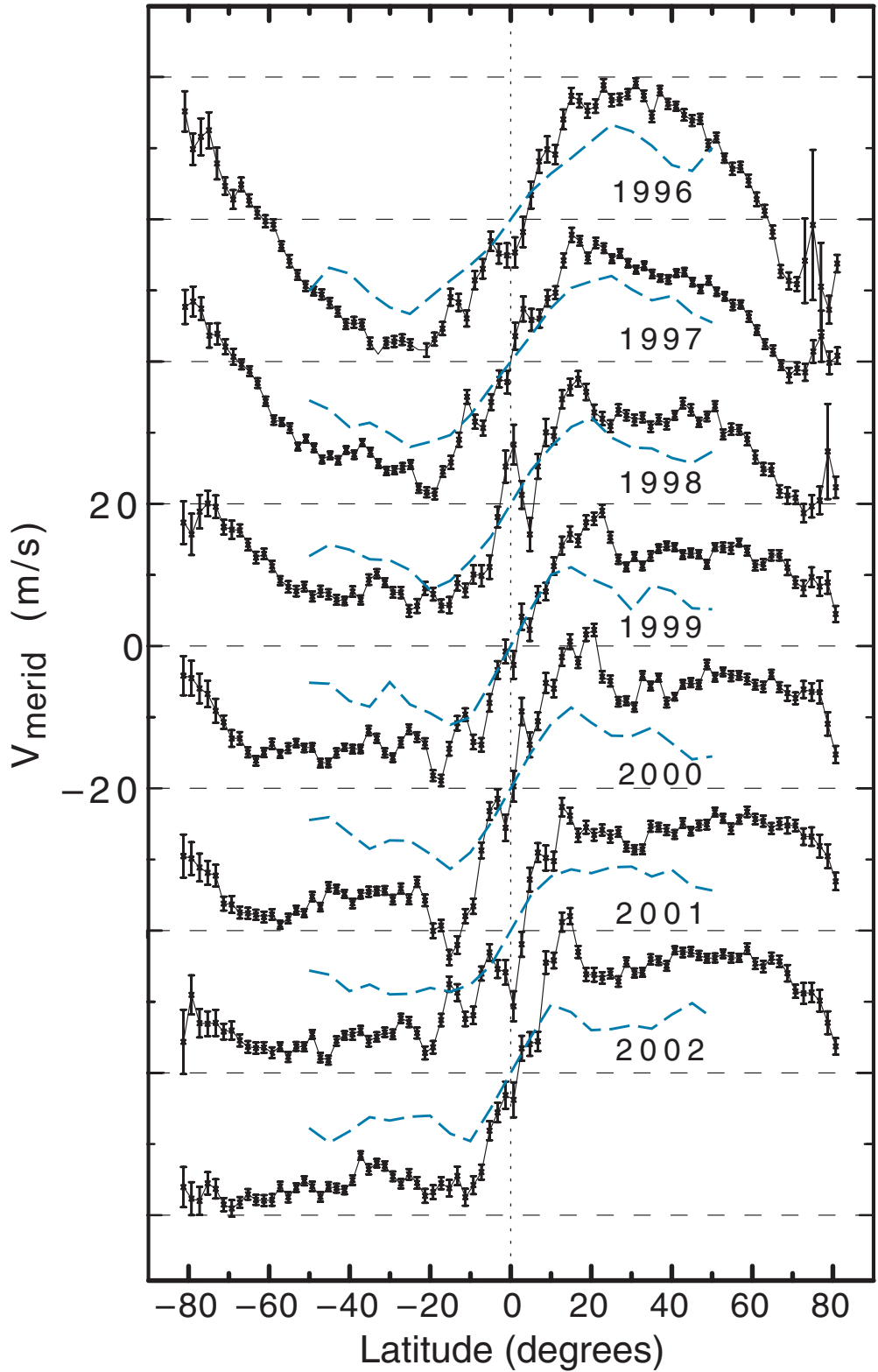}
\caption{The meridional circulation velocity derived by \citet{2008SoPh..251..241G} based on feature tracking of the surface supergranulation pattern plotted as blue solid lines.  The comparison lines with error bars are extracted from figure \ref{yearly_circ} above.  The supergranulation feature tracking lines are derived from the periods when the MDI spacecraft carries out it dynamics campaign and involve only about one month per year.  The comparison lines from this study are yearly averages so the sampled times are not identical.
\label{CompareToSG}}
\end{figure}

The primary results are given in figure \ref{yearly_circ}.  The time dependence seen in this figure confirms similar variations found by various methods.  The meridional circulation changes sign at high latitude for several of the years with this pattern being present for a longer period following the sunspot minimum between cycles 21 and 22 than for a similar period between cycles 22 and 23.  The average of the circulation velocity over each cycle confirms the pattern used by \citet{2010GeoRL..3714107D} in a model for the greater length of cycle 23 compared to cycle 22. This reversed cell may be an effect from the polar magnetic field which reach maximum strength at sunspot minimum.  If the reversed cell is a rebound effect, the weaker polar fields at sunspot minimum between cycles 22 and 23 compared to the similar time between cycles 21 and 22 would have caused a weaker reversed flow and required the meridional circulation to extend to higher latitude as proposed by \citet{2010GeoRL..3714107D}.

Comparison of the Doppler-shift derived meridional circulation to that derived from other patterns shows that the Doppler-shift result reported here and the ring-diagram result \citep{2002ApJ...570..855H} are in good agreement apart for a 10\%\ lower amplitude from the Doppler-shifts.  The magnetic feature-tracking results agree with each other \citep{1993SoPh..147..207K,2010Sci...327.1350H} but disagree with the results here.  The divergence of the supergranulation yields a circulation pattern about 60\%\ of that from the current Doppler-shift results.  It is evident that different markers of position on the solar surface drift at different rates and are responding to different phenomena.  The Doppler-shift is most closely related to the solar surface plasma and should provide the most direct measure of the solar surface hydrodynamic state.  The magnetic features then provide an indication of the drift of the magnetic field lines through the plasma.

The MWO synoptic program at the 150-foot tower telescope includes full measurments of three other spectral lines that probe different altitudes in the solar photosphere and low chromosphere.  An examination of these additional data sets is planned and should provide an indication of the possible presence of vertical shear in the photosphere.  The nearly continuous data sequence was interrupted for 4.5 months in late 2009 and early 2010 due to equipment failure and the lack of site access following the Station Fire in the Angeles National Forest.  The synoptic program has resumed and the data sequence described here is currently being extended.  The most recent data from 2010 continues to show the polar reversed cell that began in 2007. 

\acknowledgments
I thank John E.\ Boyden for maintaining and developing the data reduction software system and for his advice in its utilization.  I thank Steve Padilla, Larry Webster and Pam Gilman for their dedicated service as observers whose work made this analysis possible.  The synoptic program has benefited from support by NASA, NSF and ONR over many years.  Current support comes from NSF through grant AGS-0958779, NASA through grants NNX09AB12G and HMI subcontract 16165880 and NASA through grant NNX08AQ34G as a subcontract from the High Altitude Observatory.
\newpage

\clearpage



\clearpage

\clearpage

\clearpage

\clearpage

\clearpage

\clearpage

\clearpage

\clearpage

\clearpage

\clearpage

\clearpage

\end{document}